\documentclass[conference]{IEEEtran}
\IEEEoverridecommandlockouts
\usepackage{setspace}
\usepackage{graphicx}
\usepackage{amsmath}
\usepackage{amssymb}
\usepackage{booktabs}
\usepackage{url}
\usepackage{multirow}
\usepackage{multicol}
\usepackage{algpseudocode}
\usepackage{listings}
\usepackage{tablefootnote}
\usepackage[dvipsnames]{xcolor}
\usepackage{enumitem}
\usepackage{algorithm}
\usepackage{algpseudocode}
\usepackage{sidecap,wasysym,array,tabularx,caption}

\algrenewcommand{\Return}{\State\algorithmicreturn~}
\usepackage{colortbl}
\usepackage[normalem]{ulem}
\usepackage{relsize,amsfonts}
\usepackage{import}
\definecolor{mygray}{gray}{.9}
\def\BibTeX{{\rm B\kern-.05em{\sc i\kern-.025em b}\kern-.08em
    T\kern-.1667em\lower.7ex\hbox{E}\kern-.125emX}}
    
\usepackage{hyperref}
\usepackage{url}
\definecolor{mygray}{gray}{.9}

\usepackage{fancyhdr}
\begin{document}
\title{Emolysis: A Multimodal Open-Source Group Emotion Analysis and Visualization Toolkit}

\author{\IEEEauthorblockN{Shreya Ghosh*}
\IEEEauthorblockA{\textit{Curtin University}\\
Perth, Australia \\
shreya.ghosh@curtin.edu.au}
\and
\IEEEauthorblockN{Zhixi Cai*}
\IEEEauthorblockA{\textit{Monash University}\\
Melbourne, Australia \\
zhixi.cai@monash.edu}
\and
\IEEEauthorblockN{Parul Gupta}
\IEEEauthorblockA{\textit{Monash University}\\
Melbourne, Australia \\
parul@monash.edu}
\and
\IEEEauthorblockN{Garima Sharma}
\IEEEauthorblockA{\textit{Monash University}\\
Melbourne, Australia \\
garima.sharma1@monash.edu}
\and
\IEEEauthorblockN{Abhinav Dhall}
\IEEEauthorblockA{\textit{Flinders University}\\
Adelaide, Australia \\
abhinav.dhall@flinders.edu.au}
\and
\IEEEauthorblockN{Munawar Hayat}
\IEEEauthorblockA{\textit{Qualcomm}\\
San Diego, United States \\
hayat@qti.qualcomm.com}
\and 
\IEEEauthorblockN{Tom Gedeon}
\IEEEauthorblockA{\textit{Curtin University}\\
Perth, Australia \\
tom.gedeon@curtin.edu.au}
}

\setstretch{0.99}

\maketitle
\thispagestyle{fancy}

\begin{abstract}
Automatic group emotion recognition plays an important role in understanding complex human-human interaction. This paper introduces,~\textit{Emolysis}, a Python-based, standalone open-source group emotion analysis toolkit for use in different social situations upon getting consent from the users. Given any input video, Emolysis processes synchronized multimodal input and maps it to group level emotion, valence and arousal. Additionally, the toolkit supports major mobile and desktop platforms (Android, iOS, Windows). The Emolysis platform also comes with an intuitive graphical user interface that allows users to select different modalities and target persons for more fine-grained emotion analysis. Emolysis is freely available for academic research and encourages application developers to extend it to application specific environments on top of the existing system. We believe that the extension mechanism is quite straightforward. Our code models and interface are available at \href{https://github.com/ControlNet/emolysis}{https://github.com/ControlNet/emolysis}.
\end{abstract}


\def\thefootnote{*}\footnotetext{These authors contributed equally to this work}

\section{Introduction and Background}
Emotion is an essential part of human experience which impacts human cognition, perception, learning, communication, decision-making and many other daily life scenarios~\cite{dhall2015automatic,ghosh2020automatic,ghosh2018role,sharma2023graphitti}. Being social-beings, our human minds incline towards any group environment which further develops several concomitants such as interpersonal and intrapersonal emotion, bonding, interaction etc~\cite{ghosh2020automatic}. The prior work in group level emotion analysis approaches can be divided into three broad categories: \textit{bottom-up approaches}, \textit{top-down approaches} and \textit{hybrid approaches}. The \textit{bottom-up approaches} analyze the group members individually and then assess the contribution of these members towards the overall group's mood~\cite{MoodMeter}. On the other hand, the main motivation behind \textit{top-down approach} is to determine global factors and further analyze how that impacts the perception of a group's emotion~\cite{barsade1998group}. Finally, \textit{Hybrid approaches} use both holistic level and individual level information ~\cite{mou2016alone,ghosh2018automatic,sharma2021audio,sharma2019automatic}. Group level emotion estimation has potential applications in getting automatic feedback of a lecture~\cite{whitehill2014faces}, student engagement in a class~\cite{kaur2018prediction}, event detection~\cite{vandal2015event}, surveillance, image ranking from an event~\cite{dhall2015automatic}, event summarization~\cite{dhall2015automatic} etc. Despite having such a wide range of applications, there are no visualization toolkits that support interactive, multimodal, person-specific and temporal analysis of categorized emotion along with fine grained valence and arousal mapping. The main contributions are as follows:

\begin{itemize} [topsep=1pt,itemsep=0pt,partopsep=1ex,parsep=1ex,leftmargin=*]
    \item \textit{Emolysis} performs interactive, multimodal categorization of group emotion along with valence arousal analysis.
    
    \item To the best of our knowledge, we are the first to develop an interactive group emotion  toolkit that offers an OS independent, flexible and efficient framework.
    
    \item \textit{Emolysis} can be easily extendable for a wide range of human behavior analysis applications (such as event and people monitoring for engagement in lectures, mob management, group dynamics etc).
\end{itemize}

\begin{figure}[t]
    \centering
    \includegraphics[height = 3.8cm]{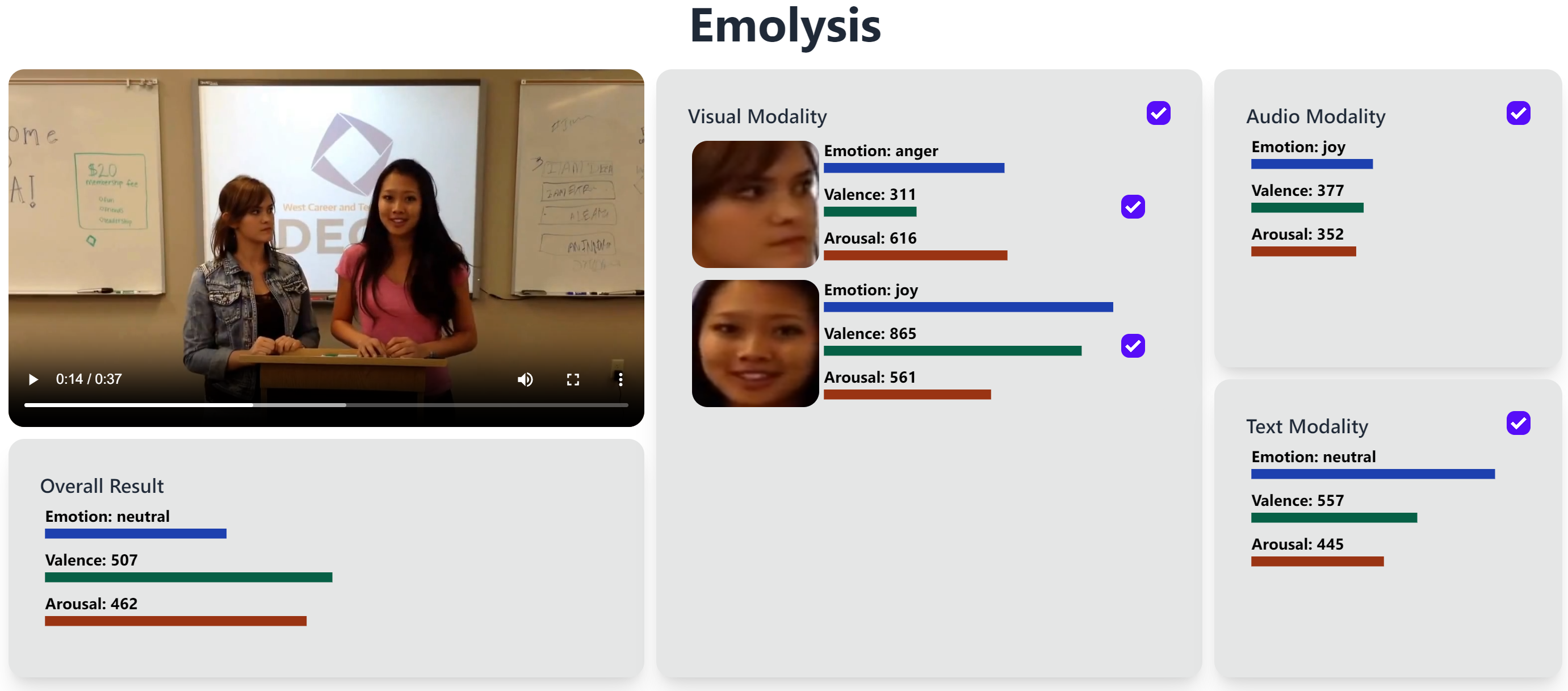}
    \caption{\small \underline{Emolysis Overview.} Emolysis is an interactive, multimodal group emotion analysis and visualization toolkit. The screenshot represents an example scenario from a video feed where the user can select a subset of people as well as the modality used.}
    \label{fig:Emolysis}
    \vspace{-7.8mm}
\end{figure}

\begin{figure}[t]
    \centering \includegraphics[height = 3.8cm]{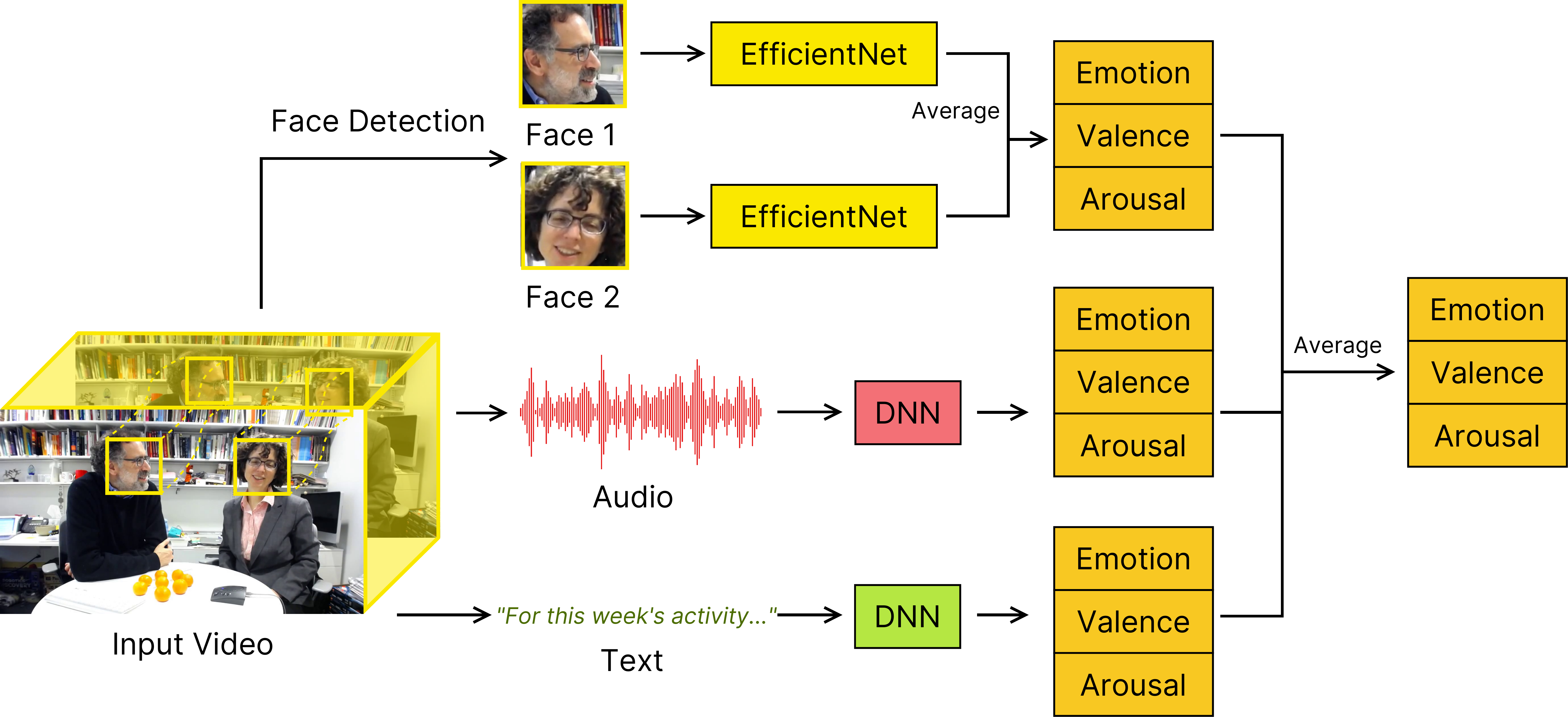}
    \caption{\small \underline{Emolysis Workflow.} The workflow of the Emolysis pipeline. Emolysis tool allows the user to interact with the interface in terms of modality as well as subset of person selection.}
    \label{fig:components}
    \vspace{-6mm}
\end{figure}


\vspace{-2mm}
\section{Emolysis: Software Description}

Emolysis is a web-based user interface designed for the analysis of any input video (See Fig.~\ref{fig:components}). 


\noindent \textbf{Input.} In the Emolysis user interface, we provide options to the user to choose a specific video file from any local/online storage path. Upon uploading the video, we process it segment-wise and provide synchronized input to the modality-specific neural networks. Additionally, we allow the users to select specific persons and modalities for more fine-grained analysis. Please note that the system currently supports two languages, English and Mandarin. 

\noindent \textbf{Visual.} The input to the visual modality is a segment of detected face frames stacked sequentially. The EfficientNet~\cite{tan2019efficientnet} based visual emotion, valence and arousal detection network is adopted from prior work~\cite{savchenko2021facial}. First, we detect and track frame-specific faces using the MTCNN face detection module~\cite{zhang2016joint}. The detected faces are then cropped and resized into $224\times 224$ images for fast computation. For the backbone visual module, we incorporate the HSE model~\cite{9815154}, which is pre-trained on AffectNet dataset~\cite{mollahosseini2017affectnet} to predict the results for each face. 

\noindent \textbf{Audio.} We first extract the synchronized audio from the uploaded video. Here, we pre-process the audio to split it into a 15s sliding window with 7.5s strides to maintain local as well as global factors. We further extract TRILL features~\cite{shor2020towards} from the input raw audio segments by TensorFlow library~\cite{abadi2015tensorflow}. We incorporate a DNN module on top of TRILL features to predict audio emotion, valence and arousal scores. Here, the main aim is to encode non-semantic information which is highly coherent with group emotion. We train the aforementioned DNN module on the ARBEE dataset~\cite{luo2020arbee}.

\noindent \textbf{Linguistic.} The input to the linguistic module is the transcribed text from the audio using the whisper text-to-speech library~\cite{radford2022robust}. Please note that the current version of Emolysis module only supports two languages i.e. English and Mandarin. For linguistic based emotion prediction, we incorporate a pre-trained RoBERTa module~\cite{liu2019roberta} followed by a three-layer DNN that maps the latent RoBERTa features to label space. Here, our aim is to encode semantic information from the conversation. For implementing the linguistic module, we train two different language specific models by fine-tuning the RoBERTa encoder along with the DNN based prediction head using PyTorch library~\cite{paszke2019pytorch}. While uploading the video, the user has the option to select the language. Based on user specific language selection, the system will utilize the corresponding linguistic model for inference. Please note that we use the CMU-MOSEI dataset~\cite{zadeh2018multimodal} to train the English model and the ARBEE dataset~\cite{10.1145/3442188.3445924} to train the Mandarin.

\noindent \textbf{Output: Emotion.} The label space of the Emolysis framework is the probability of Plutchik's eight class emotions~\cite{plutchik1980general} i.e. joy, trust, fear, surprise, sadness, anticipation, anger, and disgust along with a `no emotion or emotion not present' class. In the literature, group emotion is coarsely categorized into the valance axis of emotion i.e. positive, negative and neutral~\cite{dhall2015more,dhall2017individual}. For more fine-grained analysis, we use eight-class emotions in this aspect. To bridge the gap, we keep the label space as a multilabel emotion prediction problem as multiple emotions co-exists in complex human-human interaction.

\noindent \textbf{Output: Valence and Arousal.} The label space also consist of continuous emotion labels i.e. valence and arousal scores. 

\noindent \textbf{Interface:} \textit{1. Frontend.} We utilize the \textit{D3.js}~\cite{6064996} and \textit{Vue.js}~\cite{vuejs} frameworks for the frontend development. Users can interactively choose among visual, audio and text modalities as per their requirement and convenience. Additionally, user can track each individual face level emotion, valence and arousal and select/deselect them as per the application specific requirements. 
\noindent \textit{2. Backend.} We have utilized PyTorch~\cite{paszke2019pytorch} library based deep learning models with FastAPI~\cite{fastapi} framework for hosting the backend server. 

\vspace{-1mm}
\section{Performance}
\label{sec:label_space}

Please note that AffectNet, CMU-MOSEI, and ARBEE datasets have different label spaces. We further map these different label spaces to a common label space which investigates a more fine-grained aspect of emotion in a temporal fashion. 

\noindent \textbf{Performance Comparison.} The toolkit was tested on an independent set of group videos (i.e. from GAF~\cite{sharma2019automatic}) to verify the generalizability of algorithms used in the toolkit. It looks promising to 10 independent reviewers qualitatively. 






\vspace{-1mm}
\section{Conclusion} 
\noindent \textbf{Access.} Emolysis is freely available for academic research purposes at Github and DockerHub for easy deployment.

\noindent \textbf{Ethical Use and Privacy Concerns.}
Users are expected to respect the licenses of the deployed components used in Emolysis. During run-time, Emolysis performs the detection and tracking of faces when allowed access to any person's facial video. This could potentially breach the privacy of the subjects present in a group video. Thus, it is important that the users need to have informed consent from the subjects before using our toolkit. While developing this solution, we make sure that Emolysis does not capture or save any image or video in the file system. Moreover, all the operations are performed at run time; buffered face images are removed immediately after prediction. We use the off-the-shelf MTCNN~\cite{zhang2016joint} library for face detection and trained our models on different `in-the-wild' datasets. Thus, there could be potential bias in terms of race and culture. We will investigate and remove any such dependencies in our future versions.

To the best of our knowledge, \textit{Emolysis} is the first interactive toolkit for detecting multimodal and multiperson emotion, valence and arousal from any input videos. The proposed standalone toolkit works well across different OS platforms and is easily transferable. The toolkit also provides an interactive environment for unimodal, person-specific setups. We believe that \textit{Emolysis} has the potential to significantly impact the \textit{`empathetic' design process of upcoming devices and human computer interfaces}. 


\bibliographystyle{IEEEtran}
\bibliography{ref.bib}

\end{document}